\documentclass{aa}  

\usepackage{CJK}
\usepackage{color}
\usepackage{gensymb}
\usepackage{graphicx}
\usepackage{txfonts}
\usepackage{lipsum}
\usepackage{subcaption}         \usepackage{lscape}             \usepackage{placeins}           \usepackage{hyperref}
\usepackage{numprint}
\usepackage[normalem]{ulem}

\providecommand\figref[1]{Fig.~\ref{#1}}
\providecommand\figsref[1]{Figs.~\ref{#1}}
\providecommand\figrefalt[1]{Figure~\ref{#1}}

\providecommand\secref[1]{Sect.~\ref{#1}}

\begin{document}
\begin{CJK*}{UTF8}{gbsn}

   \title{Phase spirals induced by the gas warp}

   \author{Shuyu Wang (汪书玉)\inst{1}
        \and Anthony G. A. Brown \inst{1}
        \and Victor P. Debattista \inst{2}
        \and Tigran Khachaturyants \inst{3}
        }

   \institute{Leiden Observatory, Leiden University, Einsteinweg 55, 2333 CC Leiden, The Netherlands\\
             \email{shuyuwang.astro@gmail.com}
        \and Jeremiah Horrocks Institute, University of Lancashire, Preston PR1 2HE, UK
        \and Department of Astronomy, Shanghai Jiao Tong University, 800 Dongchuan Road, Shanghai 200240, P.R. China}

  \abstract
   {The discovery of the phase space spirals in the Solar neighborhood in \Gaia Data Release~2 has prompted various attempts to understand their origin. A source of bending waves, which has been neglected as a cause of the phase spiral, is irregular gas inflow along the warp.}
   {We aim to study whether perturbations by the gas warp could induce phase spirals. Accounting for this additional formation scenario for phase spirals could improve our current understanding of the perturbation history of the Milky Way disc. } 
   {We use two $N$-body$+$SPH (Smooth Particle Hydrodynamics) simulations of an isolated galaxy to search for, and study, warp-induced phase spirals. We study the emergence and propagation of the detected phase spirals using Fourier decomposition.  }
   {We detect strong one-armed phase spirals in the warped simulation. These phase spirals are prevalent and persist over $\sim 10$ Gyr. The morphology of these phase spirals varies with location and evolves with time. In particular, the emergence rate of the phase spiral evolves with the gas inflow at the outer disc and the bending wave amplitude, indicating that these phase spirals are a record of warp-induced bending waves. We find that these phase spirals can reach amplitudes comparable to those in the \Gaia~DR3. We only detect weak and stochastically distributed phase spirals in an unwarped control simulation. }
   {We conclude that phase spirals can be induced by the irregular gas accretion along the warp. These phase spirals occur globally and are long-lived. }


   \maketitle
\nolinenumbers

\section{Introduction}
\label{sec:intro}
\citet{Antoja_2018} discovered the `phase spiral' --- a coherent spiral pattern in the vertical phase space of the Solar neighborhood, using \Gaia~DR2 data \citep{2018A&A...616A...1G}. This one-armed phase spiral was observed in the $Z$-$V_z$ phase space map weighted by stellar number density, median radial velocity, or median azimuthal velocity. The discovery provides compelling evidence for the dynamically-perturbed state of the Milky Way disc and ongoing phase-mixing in the Solar neighborhood.

Subsequent observational work endeavored to map the phase spirals across the Galactic disc. The increased sample size and improved astrometric precision provided by \Gaia~DR3 \citep{2023A&A...674A...1G} allowed the identification of a sharper signature in an extended sample \citep{2023A&A...673A.115A}, and the mapping of phase spirals across Galactic radius and azimuth. \citet{10.1093/mnrasl/slac082} split a \Gaia~DR3 sample within a cylindrical distance $d_{\mathrm{cyl}} \leq 1$~kpc from the Sun into dynamically local groups in the azimuthal action (guiding center radius) and angle space. This division spanned a wide range in guiding center radii and revealed a two-armed phase spiral signature in the inner Galactic disc.

Various formation scenarios have been proposed for the one-armed (two-armed) phase spirals, which are typically associated with the vertical bending (breathing) waves in the Galaxy. The most prevalent hypothesis supposes that one-armed phase spirals arise from the recent passage of the Sagittarius dwarf galaxy (Sgr) \citep[e.g.,][]{10.1093/mnras/sty2378, 2019MNRAS.486.1167B, 2019MNRAS.485.3134L, 10.1093/mnras/stab704, 10.1093/mnras/stab2580, 2025A&A...700A.109A}. One-armed phase spirals may also result from the dark matter wake produced by the Large Magellanic Cloud (LMC) \citep{2023MNRAS.524..801G}. They may be induced by internal processes, such as the buckling of the Galactic bar \citep{2019A&A...622L...6K}. Additionally, one-armed phase spirals may arise from the combined effect of stochastic heating by the dark matter (DM) subhalos \citep{2023MNRAS.521..114T,2025ApJ...980...24G,10.1093/mnras/staf1331}. A slowing bar \citep{2023MNRAS.524.6331L} or transient spiral arms \citep{10.1093/mnrasl/slac082,2023ApJ...952...65B} may be responsible for the presence of two-armed phase spirals. \citet{2025MNRAS.543.2159C} proposed that two-armed phase spirals could naturally form in the presence of vertical resonances or stochastic kicks. The two-armed phase spirals may also be excited by the tidally-induced spiral arms \citep{2025A&A...700A.109A}.  \citet{2025ApJ...988..254L} suggested that the observed two-armed phase spiral signatures could be a superposition of one-armed phase spiral signatures from multiple previous perturbations. \citet{2025NewAR.10001721H} provide a comprehensive review of the proposed formation scenarios for phase spirals.

Characteristics of phase spirals, such as their degree of winding, amplitude, and dominant Fourier modes, provide clues to the past perturbation events in the Milky Way disc \citep[e.g.,][]{2023ApJ...955...74D,2025arXiv250719579W}, and the Galactic potential \citep[e.g.,][]{2021A&A...650A.124W,Guo_2024}. For example, in the single impact origin scenario, the degree of winding of the phase spirals could reveal the impact time --- the time elapsed since the impact seeded the vertical corrugations. By `rewinding' the phase spirals, \citet{2023A&A...673A.115A} and \citet{10.1093/mnras/stad908} estimated a Sagittarius impact time of approximately a few hundred Myr ago. However, these estimates did not include the effect of self-gravity, which can significantly reduce the winding rate of the phase spirals \citep{10.1093/mnras/sty3508, 10.1093/mnras/stad973}. Self-consistent models have further demonstrated that phase spirals commonly emerge with a delay time of a few hundred Myr after the most recent impact \citep[e.g.,][]{10.1093/mnras/stab704,10.1093/mnras/stab2580,2025A&A...700A.109A}, implying that earlier works may have systematically underestimated the impact time.

While the above studies have demonstrated the plausibility of various formation scenarios for phase spirals, none quantitatively reproduces the \Gaia phase spirals. For instance, the hypothesis that the Sgr passage alone could produce the \Gaia phase spirals would require a massive Sgr progenitor (on the order of $10^{10} M_{\odot}$, \citealt{2019MNRAS.486.1167B,2019MNRAS.485.3134L}), which cannot be reconciled with the observed low mass of the current day Sgr remnant ($\sim 4\times10^{8} M_{\odot}$, \citealt{10.1093/mnras/staa2114}). Similarly, \citet{2025ApJ...980...24G} suggested that while stochastic heating by multiple dark matter subhalos \citep{2023MNRAS.521..114T} could result in a significant perturbation, it fails to reach the amplitude of the \Gaia phase spirals.

Another source of perturbation is misaligned gas accretion.
\citet{10.1093/mnras/stac1491} and \citet{2024A&A...683A..47G} identified phase spirals in a cosmological environment with infalling satellite galaxies, a misaligned dark matter halo, minor dark matter subhalos, and misaligned gas accretion. \citet{2022MNRAS.512.3500K} discovered that misaligned gas accretion induces long-lived bending waves in an isolated warped galaxy simulation. These vertical corrugations may manifest as spiral patterns in phase space, although whether the warp can excite phase spirals was not explored.

In this work, we present the discovery of phase spirals in the isolated, warped galaxy simulation of \citet{2022MNRAS.512.3500K}. We propose that misaligned gas accretion along the Galactic warp could be a viable mechanism for forming phase spirals. Although we do not attempt to quantitatively reproduce the \Gaia observations, we detect warp-induced phase spirals that are qualitatively similar to the observed phase spirals.
This new formation channel might also contribute to the `everything' option, where the combined effect of multiple perturbers gives rise to the observed phase spirals \citep[see, e.g.][]{2025NewAR.10001721H,2025arXiv250719579W}. The proposed mechanism could deepen our understanding of the perturbation history of the Galactic disc.

In this work, we focus on analyzing the stellar disc, which is the main component where the phase space structure has been studied observationally. This work is organized as follows: in \secref{sec:sim}, we briefly recapitulate the simulations. In \secref{sec:method}, we describe the method for identifying the phase spirals in the simulations, which is adapted from \citet{10.1093/mnras/stac1491} and \citet{10.1093/mnras/staf1331}. In \secref{sec:result}, we present the detected phase spirals in the warped galaxy, their spatial variation and time evolution. In \secref{sec:discuss}, we discuss several characteristics of the detected phase spirals and provide a qualitative comparison between our results and the \Gaia observations. We summarize our results in \secref{sec:conclude}.

\section{Simulations}
\label{sec:sim}

We use two $N$-body$+$SPH (Smooth Particle Hydrodynamics) simulations presented in \citet{2022MNRAS.512.3500K}. Both models were evolved using the SPH code {\sc gasoline} \citep{Wadsley+2004}, with one model developing a warp through an initially misaligned gaseous corona, and the other remaining unwarped throughout its evolution.

The warped simulation of \citet{2022MNRAS.512.3500K} is produced via the method of \citet{vpd2015}, where the merging of two identical spherical Navarro-Frenk-White \citep{NFW1996} dark matter halos creates a triaxial dark matter halo whose embedded gaseous corona has an angular momentum misaligned with respect to the halo's principal axes.

Each progenitor halo has a virial mass and radius of $M_{200} = 8.7\times10^{11}M_{\odot}$ and $r_{200} = 196~$kpc at $z=0$, with a co-spatial gas corona comprising $10\%$ of its total mass. The gas is initially in pressure equilibrium with the overall potential and is given a spin parameter of $\lambda = 0.16$ \citep{Bullock+2001}, with specific angular momentum following $j \propto R$. The dark matter and gas components are each comprised of $10^6$ particles. Each gas particle is initialized with a mass of $1.4\times10^5~M_{\odot}$ and gravitational softening $\epsilon = 20~$pc, while dark matter particles have a softening length of $\epsilon = 100~$pc and a mass that is radially dependent: $10^6~M_{\odot}$ and $3.6\times10^6~M_{\odot}$ within and beyond $200~$kpc, respectively.

After the halos merge, this setup produces a dark matter halo with $r_{200} = 238~$kpc and $M_{200} = 1.6\times10^{12}M_{\odot}$. The hot gas corona has a spin parameter of $\lambda = 0.11$. At this stage, radiative cooling, star formation, and stellar feedback are activated, with the latter being represented with the blast-wave prescriptions of \citet{Stinson+2006}. Gas is converted into stars with 10\% efficiency when its density $n > 1$~cm$^{-3}$, its temperature $T < 15{,}000$~K, and it is part of a convergent flow. Each star particle forms with an initial mass $4.6 \times 10^4M_{\odot}$ and gravitational softening $\epsilon = 20~$pc. Star particles are represented by an entire stellar population with a Miller-Scalo \citep{MillerScalo1979} initial mass function. The evolution of star particles includes asymptotic giant branch (AGB) stellar winds and feedback from Type II and Type Ia supernovae, with their energy injected into the interstellar medium (ISM). Each supernova releases $10^{50}$ erg into the ISM.

The unwarped simulation corresponds to the M1\_c\_b model described in \citet{Karl21+}. It adopts nearly identical initial conditions to one of the progenitor halos in the warped model, but with a lower gas spin parameter, $\lambda=0.065$, yielding an initially axisymmetric disc. The feedback implementation again follows \citet{Stinson+2006}, but here the supernova energy is set to $4\times10^{50}$~erg and the star formation efficiency is reduced to $5\%$. The gravitational softening for gas particles is $50$~pc.

Both simulations were evolved for $12$~Gyr, providing complementary warped and unwarped disc systems that allowed direct comparison of the vertical structure and bending wave evolution. \citet{2022MNRAS.512.3500K} demonstrated that both simulations exhibit slow retrograde bending waves, with the amplitudes being larger in the warped model. Fast prograde bending waves were only present in the warped model. In the unwarped model, these prograde modes remained weak, consistent with expectations that they dissipate rapidly via winding. In contrast, the warped simulation maintained substantially stronger prograde waves throughout its evolution, indicating that irregular gas accretion along the warp acted as a sustained driver of vertical perturbations.

Both simulations develop a bar and multiple spiral arms. \citet{2022MNRAS.512.3500K} and \citet{10.1093/mnrasl/slac112} provided detailed analysis of the (relative) strengths, pattern speeds and time evolution of their corresponding $m = 2$ density waves.

\section{Identifying the phase spirals}
\label{sec:method}
\begin{figure}
  \resizebox{\hsize}{!}{\includegraphics{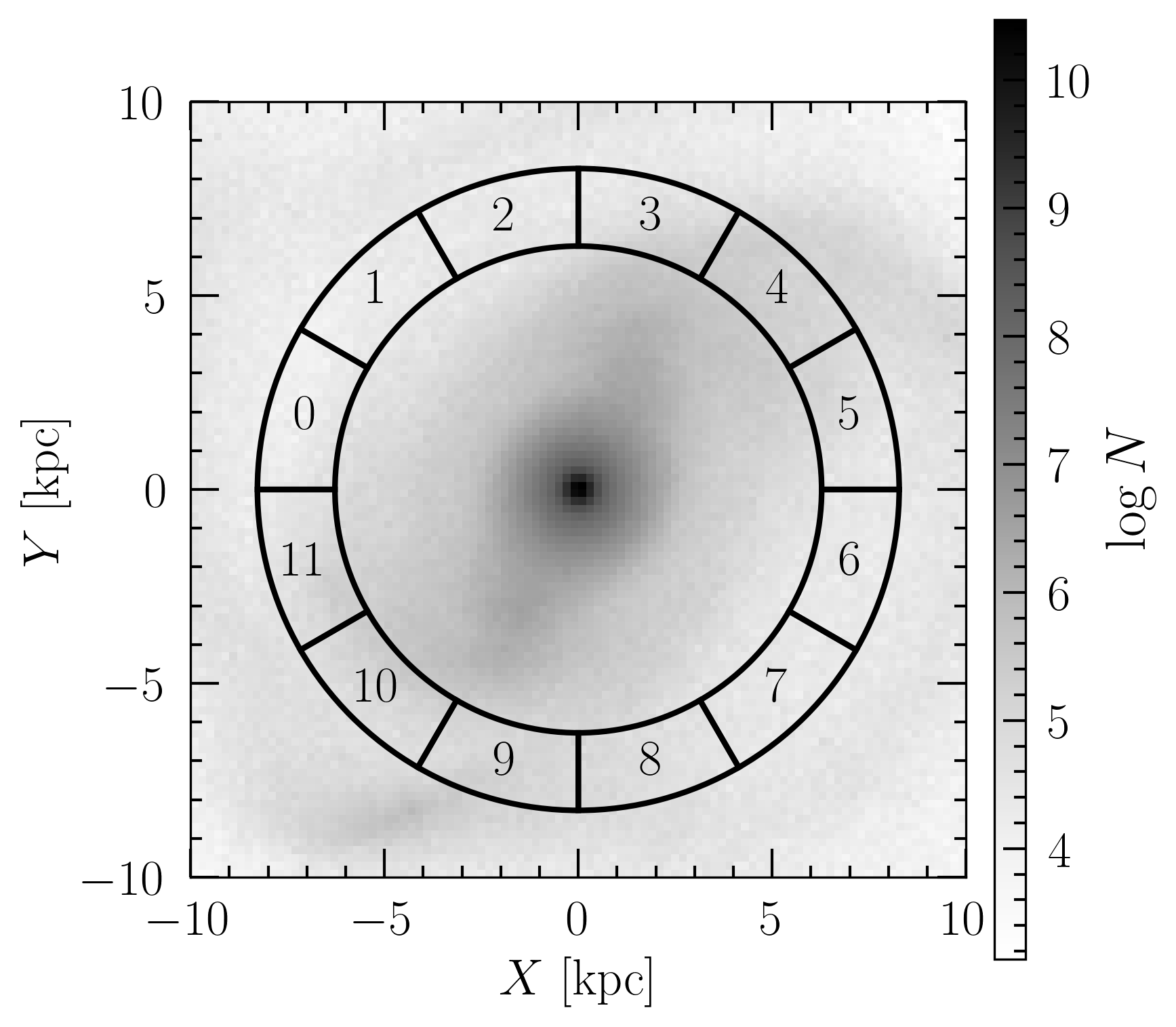}}
  \caption{The 12 regions examined for the presence of phase spirals projected onto a logarithmic star count map of the simulated warped galaxy at $8.7$~Gyr. The regions are equally spaced by $30\degree$ in azimuth with radius between $R = 6.277$ and $8.277$~kpc. The regions are labeled from 0 to 11 in a clockwise direction. The galaxy rotates counter-clockwise.}
  \label{fig:12regions}
\end{figure}

\begin{figure*}
\sidecaption
  \includegraphics[width=12cm]{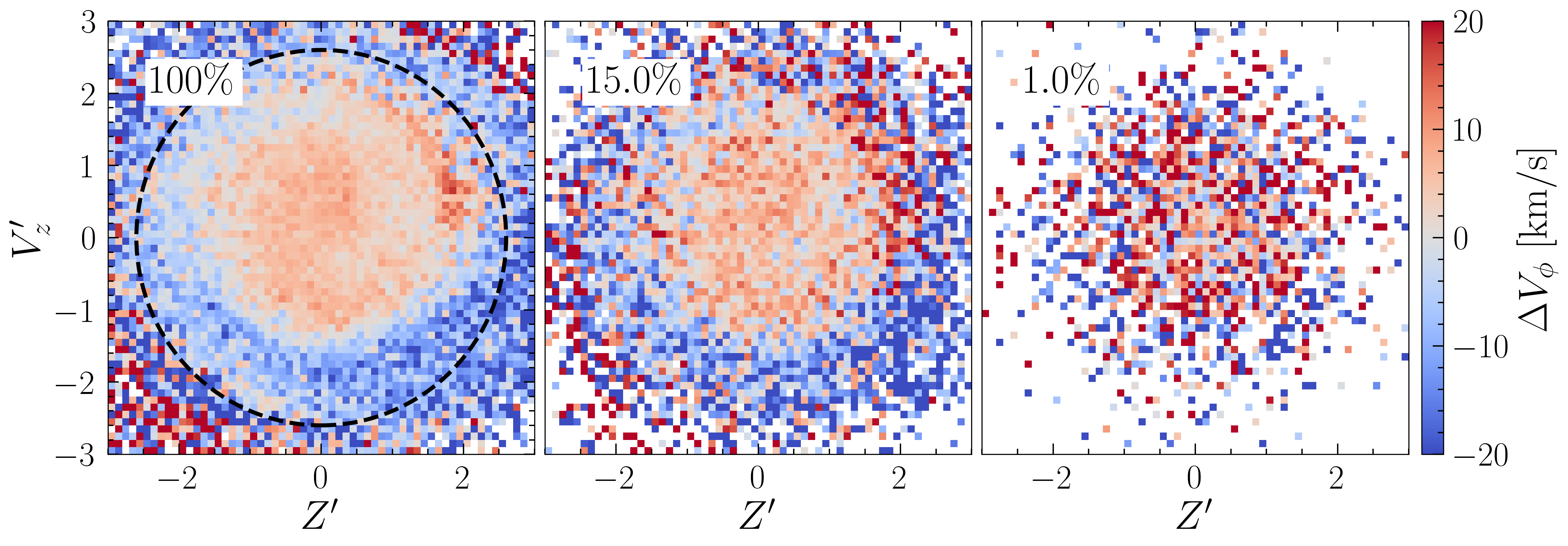}
    \caption{An example phase spiral observed with varying mass resolution in the Solar neighborhood ($d_{\mathrm{cyl}} \leq 0.8$~kpc from the Sun) of a high-resolution pure N-body simulation \citep{2025A&A...700A.109A} at $1.516$~Gyr, color-coded by residual azimuthal velocity, $\Delta V_{\phi}$. The panels show the full sample (left), and randomly selected $15\%$ (middle) and $1\%$ (right) subsamples of the full sample of $\numprint{266978}$ star particles. The black circle in the left panel indicates the cutoff at phase space radius $\tilde{R} = 2.6$. The phase spiral is less prominent in the middle panel and undetectable in the right panel.}
    \label{fig:resolution}
\end{figure*}

After centering the galaxy and aligning the disc with the $X$-$Y$ plane as described in \citet{2022MNRAS.512.3500K}, we select disc star particles with $Z \in [-2.5, 2.5]$~kpc. In order to track the evolution of phase space near the Solar radius, we divide an annulus with galactocentric radius from $R = 6.277$ to $8.277$~kpc into 12 regions equally separated by $30\degree$ in azimuth, where $R = 8.277$~kpc is the Solar radius \citep{2022A&A...657L..12G}. The galaxy is oriented such that it rotates counter-clockwise (positive sense of rotation in the right-handed $XYZ$ coordinate system). \figrefalt{fig:12regions} presents a stellar number map of the warped simulation at $8.7$~Gyr. We show the division of the 12 regions superposed on the map. The regions are labeled from 0 to 11 in a clockwise direction.

To identify and characterize the phase spirals in the simulation with a standardized method, we adapt the phase spiral finder algorithm from \citet{10.1093/mnras/stac1491}. For each region at the Solar radius (\figref{fig:12regions}), we perform a Fourier decomposition of the stellar distribution in the vertical phase space, $Z$-$V_z$. 
First, the phase space coordinates $Z$ and $V_z$ are normalized by their respective dispersions $\sigma_{Z}$ and $\sigma_{V_z}$. The normalized phase space coordinates are denoted by $Z^\prime$ and $V_z^\prime$. The dispersion $\sigma_Z$ or $\sigma_{V_z}$ is calculated using the robust scatter estimate from \citet{2016A&A...595A...4L}. 
The normalized histogram is then weighted by residual azimuthal velocity, $\Delta V_{\phi} = V_{\phi} - \mathrm{median}(V_{\phi})$, where $\mathrm{median}(V_{\phi})$ is calculated from the entire phase space histogram.\footnote{We focus on the histogram weighted by median azimuthal velocity, as the histogram weighted by median radial velocity $V_R$ is prone to being dominated by the quadrupole signatures at the outer phase space radii.}. To avoid the influence of outliers on the decomposition, we further select star particles in a region near the center of the histogram $(Z^\prime, V_z^\prime) = (0, 0)$ with phase space radius $\tilde{R} \leq 2.6$, where $\tilde{R}(Z^\prime, V_z^\prime)$ is the Euclidean distance, $\tilde{R} = [(Z^\prime)^2 + (V_z^\prime) ^2]^{1/2}$. We divide the region into $N = 20$ equal-width concentric annuli. The weighted density distribution within each annulus is decomposed into a combination of Fourier modes with $m = 0,\ldots,6$: 

\begin{equation}
    \Delta V_{\phi}(\tilde{R}, \tilde{\phi}) \approx \sum_{m=0}^6 A_m(\tilde{R})\sin(m\tilde{\phi}-P_m(\tilde{R})),
    \label{Eq:Fourier}
\end{equation}
where $\tilde{\phi}(Z^\prime, V_z^\prime)$ is calculated as $\tilde{\phi} = \mathrm{arctan2}(V_z^\prime, Z^\prime)$ and $P_m$ is a phase that varies with $\tilde{R}$. Following \citet{10.1093/mnras/stac1491} and \citet{10.1093/mnras/staf1331}, we apply the following constraints to quantitatively identify a significant $m = 1$ phase spiral: $A_1/A_0$ is higher than any $A_m/A_0$ for $m \geq 2$; $P_1$ monotonically increases with $\tilde{R}$; the scatter around a linear fit for the $P_1$-$\tilde{R}$ profile is lower than 1. We refer to the median $m=1$ Fourier amplitude $A_1$ as the strength of the phase spiral. If any of the above conditions were not met, we set $A_1 = 0$.

The Fourier decomposition method may result in false positive or false negative detections. In this work we present the conservative results by requiring that the Signal-to-Noise ratio (SNR) $A_1/\sigma{(A_1)} \geq 3$, where we bootstrap the sample of star particles within each annulus $200$ times and estimate $A_1$ and ${\sigma{(A_1)}}$ by taking the average and standard deviation of the bootstrapping results. Examples of the application of our Fourier decomposition algorithm are provided in App.~\ref{sec:fourier}.

In the phase space analysis, we adopt relatively large spatial bins due to the relatively low mass resolution of the simulations in \citet{2022MNRAS.512.3500K} (a few million star particles) compared to pure N-body simulations by e.g., \citet{10.1093/mnras/stab2580} and \citet{2025A&A...700A.109A} (more than $200$ million disc star particles). \figrefalt{fig:resolution} provides an example phase spiral with varying mass resolution in the Solar neighborhood ($d_{\mathrm{cyl}} \leq 0.8$~kpc from the Sun) for a high resolution pure N-body simulation \citep{2025A&A...700A.109A} at $1.516$~Gyr, color-coded by residual azimuthal velocity, $\Delta V_{\phi}$. The panels contain the full sample (left), randomly selected $15\%$ (middle) and $1\%$ (right) of the sample of $\numprint{266978}$ star particles, respectively. Fourier decomposition of the phase space reveals that $A_1 = 1.13\pm 0.14 \kms$, $1.39\pm 0.39 \kms$, and $3.47\pm 1.21 \kms$ for each panel, corresponding to a strong, less prominent and non-detected phase spiral. \figrefalt{fig:resolution} demonstrates how mass resolution can affect the detectability of phase spirals. We stress that our results thus describe phase spirals averaged over relatively large regions in the disc of the simulated galaxy.

\section{Results}
\label{sec:result}
In this section, we present the phase spirals detected in the warped and unwarped simulations from \citet{2022MNRAS.512.3500K}. We show examples of the phase spiral detection in the vertical phase space and study the time evolution of the occurrence of phase spirals in the 12 regions defined in \figref{fig:12regions}.

\subsection{Phase spirals in the warped galaxy} 
\label{sec:warped}

\begin{figure*}
\centering
   \includegraphics[width=17cm]{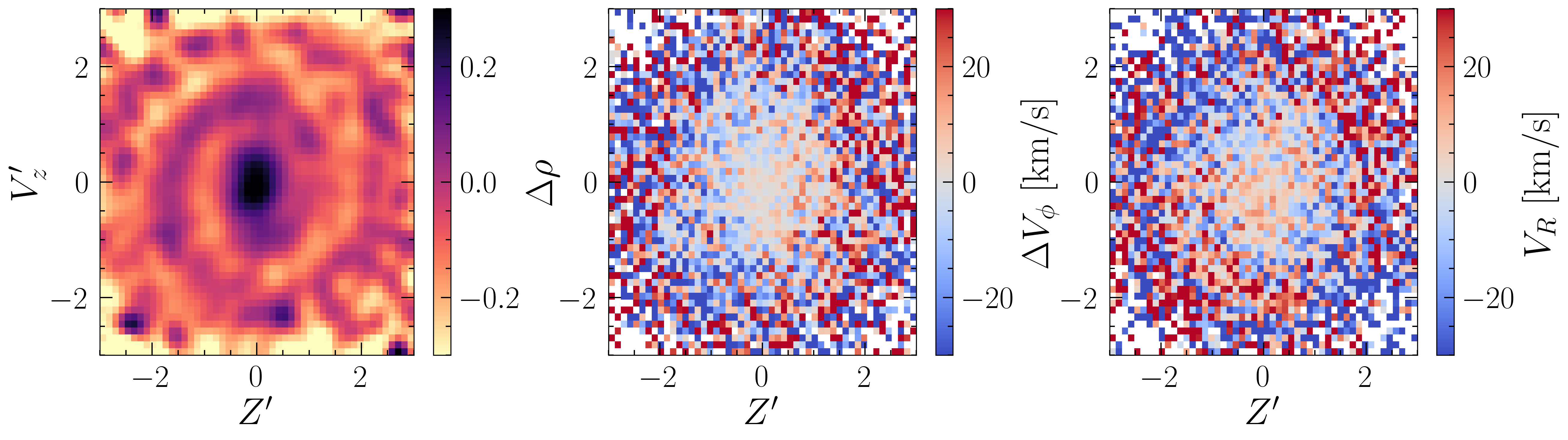}
    \caption{The phase spiral in region~4 of the warped galaxy at $8.7$~Gyr, color-coded by density contrast ($\Delta \rho$), residual azimuthal velocity ($\Delta V_{\phi}$) and median radial velocity ($V_R$), respectively.}
    \label{fig:vzz_dens_vphi_vr}
\end{figure*}

\begin{figure*}
\sidecaption
  \includegraphics[width=12cm]{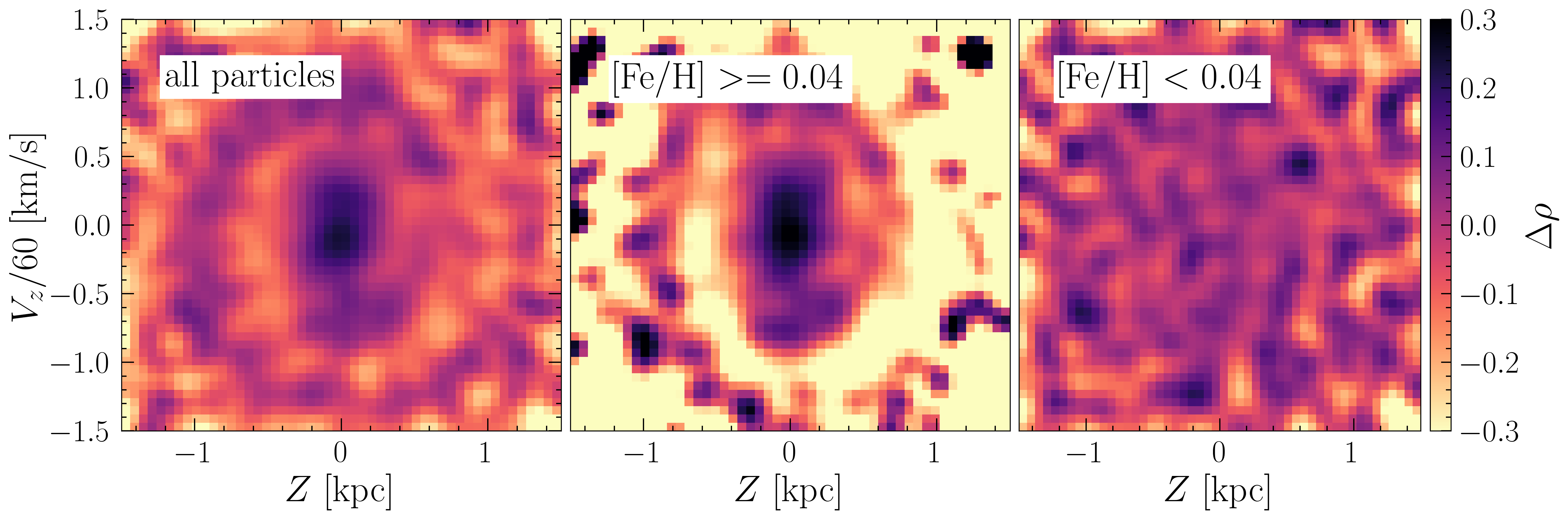}
  \caption{The phase spiral in region~4 of the warped galaxy at $8.7$~Gyr, color-coded by density contrast (left), separated into a metal-rich (middle) and a metal-poor (right) subsample at the median metallicity, $\mathrm{median([Fe / H])} = 0.04$. We scale the axes using $Z / \mathrm{kpc}$ and $V_z / 60 \kms$.}
  \label{fig:feh}
\end{figure*}

In \Gaia~DR2 and DR3 data \citep{2018A&A...616A...1G,2023A&A...674A...1G}, phase spirals were observed in the $Z$-$V_z$ maps weighted by density contrast, median radial and azimuthal velocities \citep{Antoja_2018, 2023A&A...673A.115A}, as well as by stellar metallicity \citep{2023A&A...674A..38G}. These phase space maps indicate that the vertical and in-plane velocities of the stars in the local volume are strongly correlated, a feature called `kinematic interlocking' by \citet{10.1093/mnras/staf1331}. 

As an example, we present a one-armed phase spiral detected in the warped simulation at $8.7$~Gyr. \figrefalt{fig:vzz_dens_vphi_vr} shows the phase spiral in region~4 (see \figref{fig:12regions}) of the warped galaxy, weighted by density contrast ($\Delta \rho$), residual azimuthal velocity ($\Delta V_{\phi}$) and median radial velocity ($V_R$), respectively. The region contains $\numprint{36584}$ star particles. The normalized phase space coordinates $Z^\prime$ and $V_z^\prime$ span $[-3, 3]$. The density contrast is computed as $\Delta \rho = H_1 / H_2 - 1$, where $H_1$ and $H_2$ are the stellar number counts $N$ convolved with Gaussian filters with $\sigma_1 = 1.5$ and $\sigma_2 = 3$, respectively. The phase spirals weighted by $\Delta V_{\phi}$ and $V_R$ show kinematic interlocking, consistent with the \Gaia observation.

In \figref{fig:feh}, we present the phase spiral detected in the same region for the entire sample (left), separated into a metal-rich (middle) and a metal-poor subsample (right) at the median metallicity, $\mathrm{median([Fe / H])} = 0.04$. For direct comparison with the observations, we scale the axes using $Z / \mathrm{kpc}$ and $V_z / 60 \kms$. We focus on the inner part of the phase space diagram to avoid visual artifacts that could be induced by noise at large phase space radii. \figrefalt{fig:feh} demonstrates that the inner part of the phase spiral is more prominent for the metal-rich population compared to the metal-poor population, in agreement with observations \citep{2019MNRAS.486.1167B}.

Following \citep{10.1093/mnrasl/slac082}, we bin star particles by the guiding center radius, $R_g$ and the azimuthal phase angle, $\theta_{\phi}$ to better study the spatial variation of the detected phase spirals. We select star particles with $R \in [3, 13]$~kpc, and further constrain the sample to star particles with low radial action, $J_R<\mathrm{median}(J_R)$, since they exhibit stronger bending waves \citep{2022MNRAS.512.3500K}. The action and angle variables in the analysis are derived as follows: we construct axisymmetric potential models from a simulation snapshot using \code{agama}\footnote{https://github.com/GalacticDynamics-Oxford/Agama} \citep{2019MNRAS.482.1525V}. We approximate the potential of the DM halo and hot gas ($T > \numprint{50000}$~K) using multipole expansions and that of the star particles and cold gas using an azimuthal harmonic expansion. We compute the action and angle variables using the St\"{a}ckel Fudge method \citep{10.1111/j.1365-2966.2012.21757.x} implemented in \code{agama}.

\begin{figure*}
\centering
   \includegraphics[width=17cm]{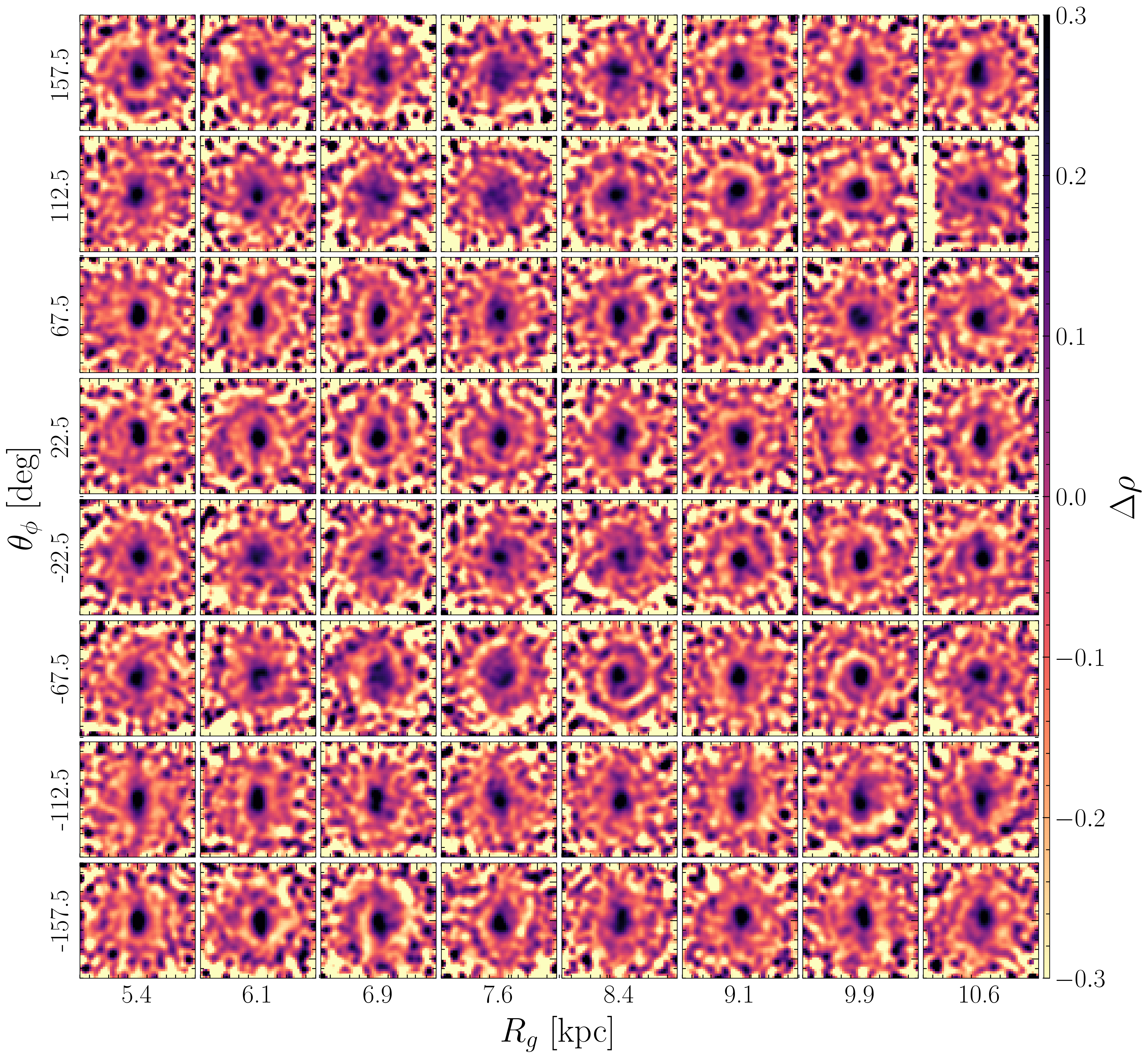}
     \caption{The phase spirals in the $R_g\text{-}\theta_{\phi}$ plane for star particles with $J_R<\mathrm{median}(J_R)$ in the warped galaxy at $8.7$~Gyr, color-coded by density contrast. Each normalized phase space coordinate covers the range $[-3, 3]$.}     \label{fig:rg_thetaphi_64_1450}
\end{figure*}
\figrefalt{fig:rg_thetaphi_64_1450} shows the resulting phase space maps in the warped galaxy at $8.7$~Gyr, color-coded by density contrast. Each normalized phase space coordinate $Z^\prime$ or $V_z^\prime$ covers the range $[-3, 3]$. 
\figrefalt{fig:rg_thetaphi_64_1450} provides a strikingly clear view of the strong phase spirals across the galaxy. A majority of these phase spirals are one-armed, though some exhibit irregular shapes, indicative of a superposition of spirals. The morphology of the phase spirals varies with location. Interestingly, their degree of winding does not vary monotonically with guiding center radius, as one would expect from an idealized model with a single impact time and no self-gravity \citep{2025arXiv250719579W}. Indeed, phase spirals appear to be most tightly wound at $R_g \in [6.9, 9.1]$~kpc. The detected phase spirals do not show a clear `rotation' with respect to the azimuthal phase angle (see \citealt{2023A&A...678A..46A}).

Note that binning star particles in the $R_g$ versus $\theta_{\phi}$ space may introduce biases in the radial action and phase angle, $\theta_R$ \citep[e.g.,][]{10.1093/mnrasl/slac082,2023ApJ...955...74D}. The axisymmetric potential approximation may also introduce biases in the radial action \citep{2025MNRAS.537.1620D}.

\subsection{Phase spiral incidence} 
\label{sec:time_incidence}
\begin{figure}
  \resizebox{8.5cm}{!}{\includegraphics{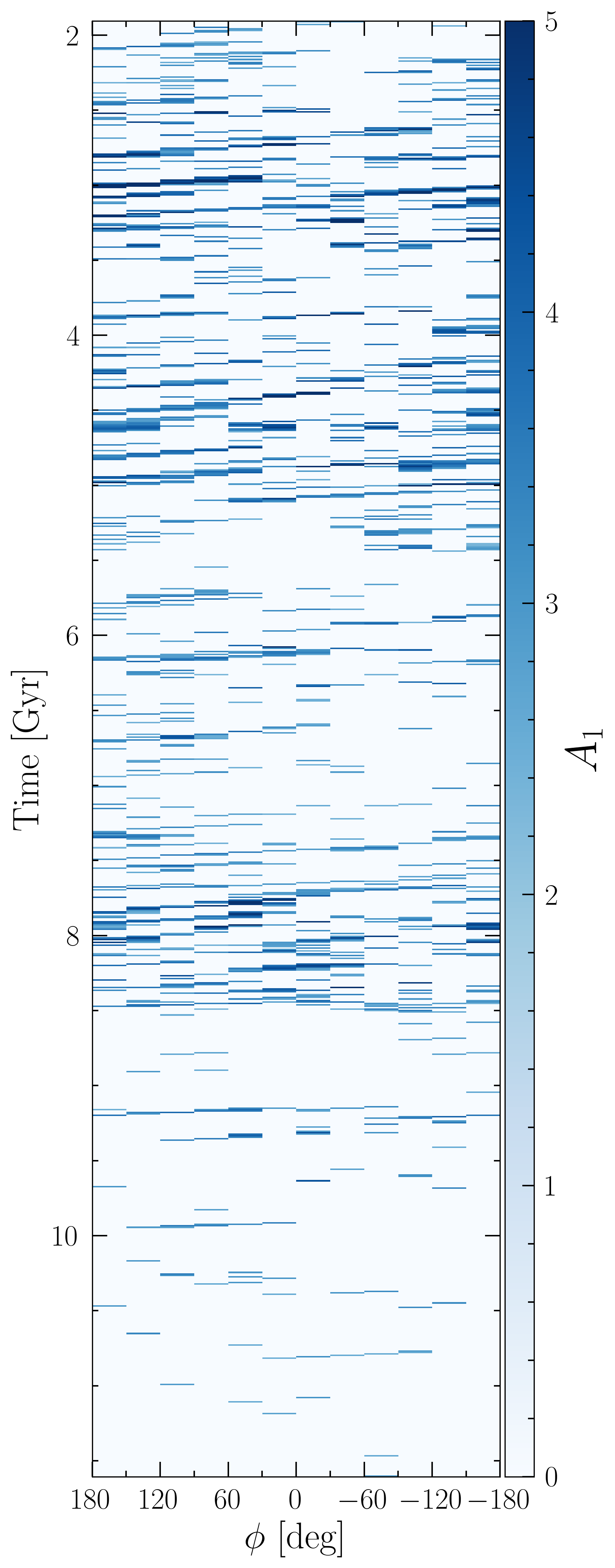}}
  \caption{Incidence map of the $m = 1$ phase spirals weighted by residual $V_{\phi}$ at the 12 regions at $R\in [6.277,8.277]$~kpc (\figref{fig:12regions}) in the warped galaxy, from $1.91$ to $11.61$~Gyr. The horizontal axis denotes the azimuth of the region, while the vertical axis denotes the time of the snapshot. The galaxy rotates from right to left along the horizontal axis.}
  \label{fig:incidence_map}
\end{figure}

Following \citet{10.1093/mnras/stac1491} and \citet{10.1093/mnras/staf1331}, we further study the spatial and time variation of the emergence and strength of the one-armed phase spirals using incidence maps. An incidence map illustrates the occurrence of phase spirals as a function of time and position within the galaxy. The map is an array of cells, where each row corresponds to the 12 regions (\figref{fig:12regions}) in a particular snapshot, and each column corresponds to different times, allowing us to track the evolution of the phase spirals. Non-empty cells are color-coded by the strength of the phase spiral $A_1$.

In \figref{fig:incidence_map}, we present the incidence map from $1.91$ to $11.61$~Gyr for our chosen annulus. The figure shows that at early times ($1.91$ to $8.51$~Gyr) one-armed phase spirals are prevalent in the warped galaxy. The average phase spiral strength over the 12 regions peaks at approximately $2.8$~Gyr, $4.4$~Gyr and $8$~Gyr. At late times ($8.51$ to $11.61$~Gyr) the one-armed phase spirals are evidently weaker and less widespread, with only a few isolated detections. In general the phase spiral incidence varies coherently with azimuth and time, manifesting as upward slopes in the incidence map. The most prominent slope has a period of $\sim230$~Myr. These slopes highlight the azimuthal propagation of the bending waves in the warped simulation.

\subsection{Time evolution of the phase spiral incidence rate}
\begin{figure*}
\sidecaption
  \includegraphics[width=12cm]{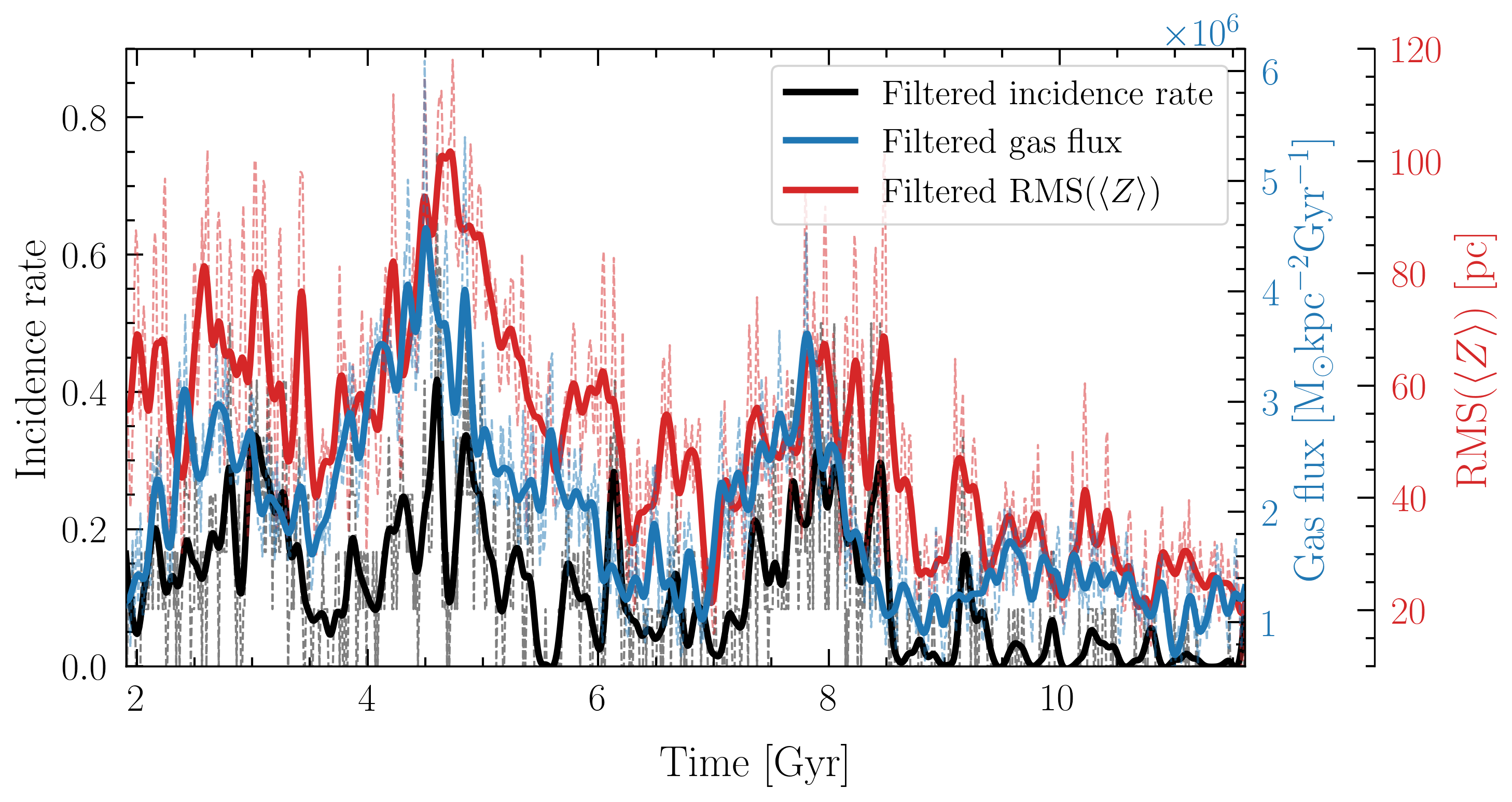}
     \caption{The time evolution of the phase spiral incidence rate at $R\in [6.277,8.277]$~kpc from $1.91$ to $11.61$~Gyr, in the warped galaxy (dashed black line) versus the gas inflow through a spherical shell at 15~kpc (dashed blue line) and the amplitude of the bending waves in the same region (dashed red line). The Butterworth filter implemented in \code{scipy} with a period of $0.175$~Gyr is applied to smooth the signals, where the filtered signals are denoted by thick solid lines. }
 \label{fig:incidence_rate_gas_inflow_flux}
\end{figure*}

We use the incidence rate of the one-armed phase spirals to study their time evolution and to compare it with that of the gas inflow and bending wave amplitude. The (dimensionless) phase spiral incidence rate at a specific time is calculated by counting the number of regions with a phase spiral detection and dividing by the total number of regions (12). \figrefalt{fig:incidence_rate_gas_inflow_flux} illustrates the time evolution of the phase spiral incidence rate in the warped galaxy versus the gas inflow and bending wave amplitude, from $1.91$ to $11.61$~Gyr. The gas inflow is calculated as the inward mass flux of cold gas ($T < \numprint{50000}$~K) through a spherical shell with radius $R = 15$~kpc and thickness $\delta R = 0.2$~kpc \citep{2022MNRAS.512.3500K}. The bending wave amplitude is estimated as the root mean square (RMS) of the mean vertical position of the 12 regions, $\mathrm{RMS}(\langle Z\rangle)$. In order to reduce high frequency noise, we apply the Butterworth filter implemented in \code{scipy} with a cutoff period of $0.175$~Gyr.

At early times ($1.91$ to $8.51$~Gyr), the phase spiral incidence rate evolves with the bending wave amplitude and gas inflow, displaying a similar triple-peaked behavior. A cross-correlation test between the filtered incidence rate and gas flux reveals a delay time of approximately $281$~Myr. The delay in the formation of phase spirals has been identified in previous simulations with externally-induced phase spirals \citep[see, e.g.,][]{10.1093/mnras/stab704, 10.1093/mnras/stab2580, 2025A&A...700A.109A}, and is consistent with the delay between the bending wave amplitude and the gas flux (approximately $250$~Myr at the Solar radius, \citealt{2022MNRAS.512.3500K}). A similar test shows that the phase spiral incidence rate evolves with the bending wave amplitude, displaying no significant delay.
The high gas inflow at early times induces strong bending waves, resulting in high phase spiral incidence rates which average $\sim 0.15$.
At late times ($8.51$ to $11.61$~Gyr), the disc thickens and the gas inflow at the outer disc decreases, resulting in dampened bending waves and low phase spiral incidence rates.

The warped galaxy forms a bar at $\sim 7.57$~Gyr that does not buckle and one-armed phase spirals have been detected well before the formation of the bar. The detected one-armed phase spirals in the warped galaxy are thus not induced by bar buckling \citep{2019A&A...622L...6K} in this simulation, but by the irregular gas inflow along the warp. In particular, by continuous excitation, the warp is able to sustain the phase spirals over the entire evolution of the simulation ($\sim 10$~Gyr). 

\subsection{Phase spirals in the unwarped galaxy} 
\label{sec:unwarped}
\begin{figure}
  \resizebox{7cm}{!}{\includegraphics{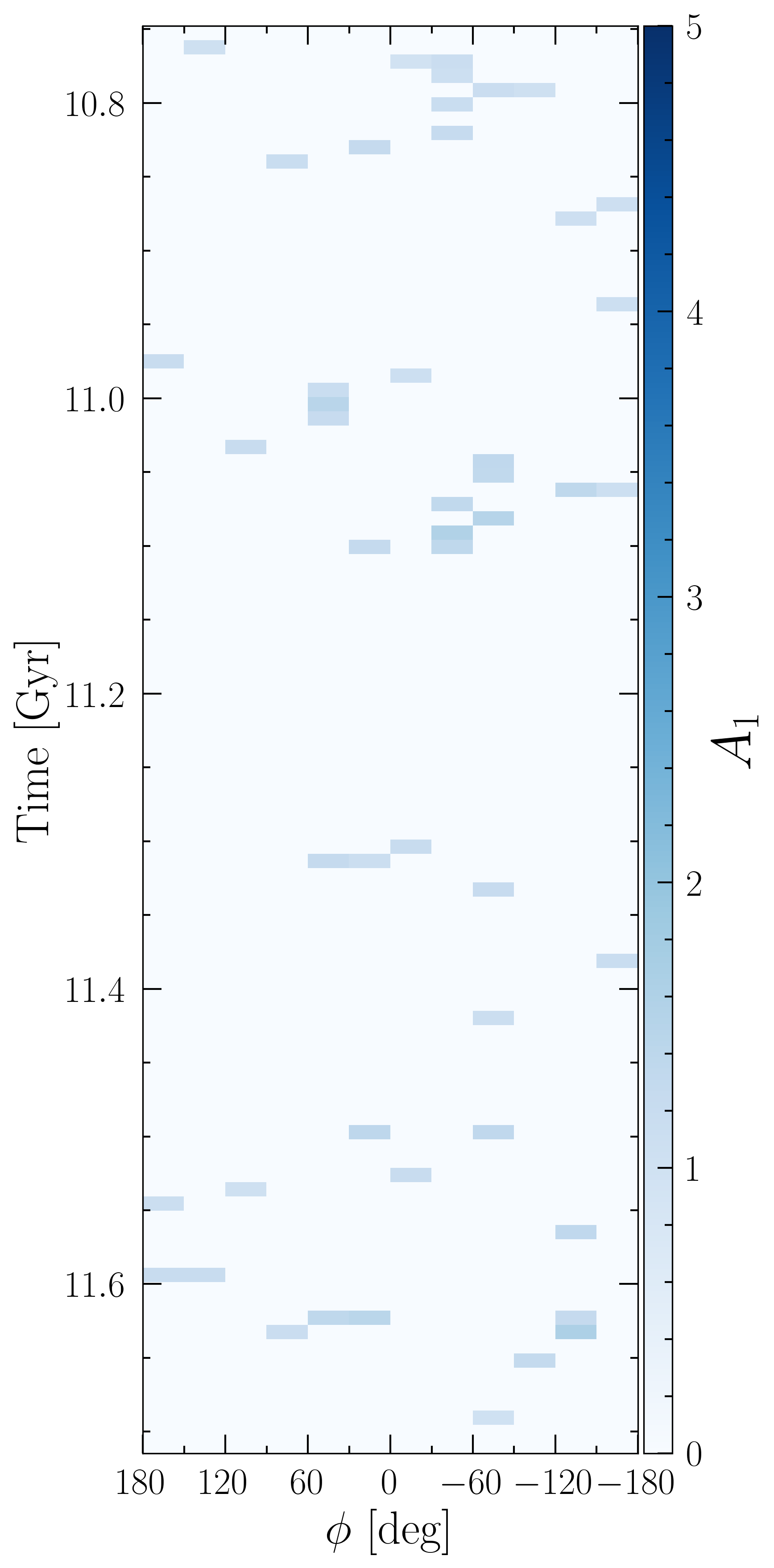}}
  \caption{Incidence map of the $m = 1$ phase spirals weighted by $\Delta V_{\phi}$ at 12 regions in the unwarped galaxy, from $10.75$ to $11.71$~Gyr.}
  \label{fig:incidence_map_unwarped}
\end{figure}

\begin{figure}
  \resizebox{\hsize}{!}{\includegraphics{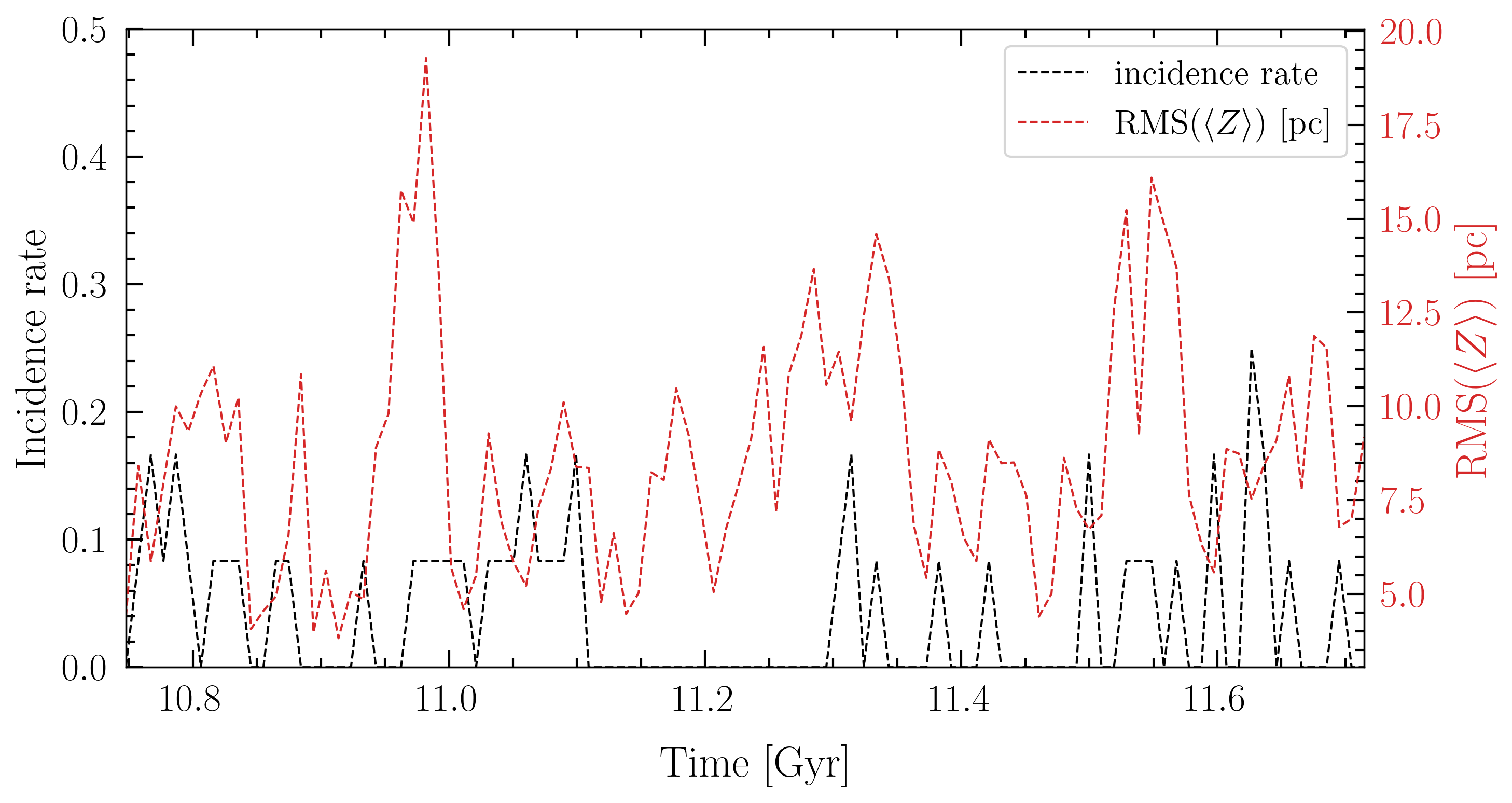}}
  \caption{The time evolution of the phase spiral incidence rate at $R\in [6.277,8.277]$~kpc from $10.75$ to $11.71$~Gyr in the unwarped galaxy (dashed black line) versus the amplitude of the bending waves in the same region (dashed red line).}
  \label{fig:incidence_rate_bending_unwarped}
\end{figure}
In this subsection, we show the phase spirals detected in the unwarped control simulation. We study disc star particles with $Z\in [-1, 1]$~kpc. \figrefalt{fig:incidence_map_unwarped} shows the incidence map of the one-armed phase spirals in the unwarped galaxy weighted by $\Delta V_{\phi}$, from $10.75$ to $11.71$~Gyr. These phase spirals are weak and appear intermittently. \figrefalt{fig:incidence_rate_bending_unwarped} illustrates the time evolution of the incidence rate of phase spirals in the unwarped galaxy versus that of the bending wave amplitude. The bending waves in the unwarped simulation are significantly weaker than those in the warped simulation \citep{2022MNRAS.512.3500K}. The time evolution of the phase spiral incidence rate does not show strong correlation with the bending wave amplitude.

Compared to those in the warped galaxy (see \figsref{fig:incidence_map} and \ref{fig:incidence_rate_gas_inflow_flux}), phase spirals in the unwarped galaxy are weaker, randomly-distributed, and only weakly-correlated with the bending waves. Despite the higher mass resolution of the unwarped simulation ($\sim 11$ million star particles at $\sim 11$~Gyr), no strong phase spirals have been detected. The phase spirals in the unwarped simulation may be induced by stochastic heating \citep{2023MNRAS.521..114T,10.1093/mnras/staf1331}, or the weak and spontaneously-generated bending waves \citep{2017MNRAS.472.2751C,2022MNRAS.512.3500K}.

\section{Discussion}
\label{sec:discuss}
\subsection{Comparison with \Gaia data}
\begin{figure}
  \resizebox{\hsize}{!}{\includegraphics{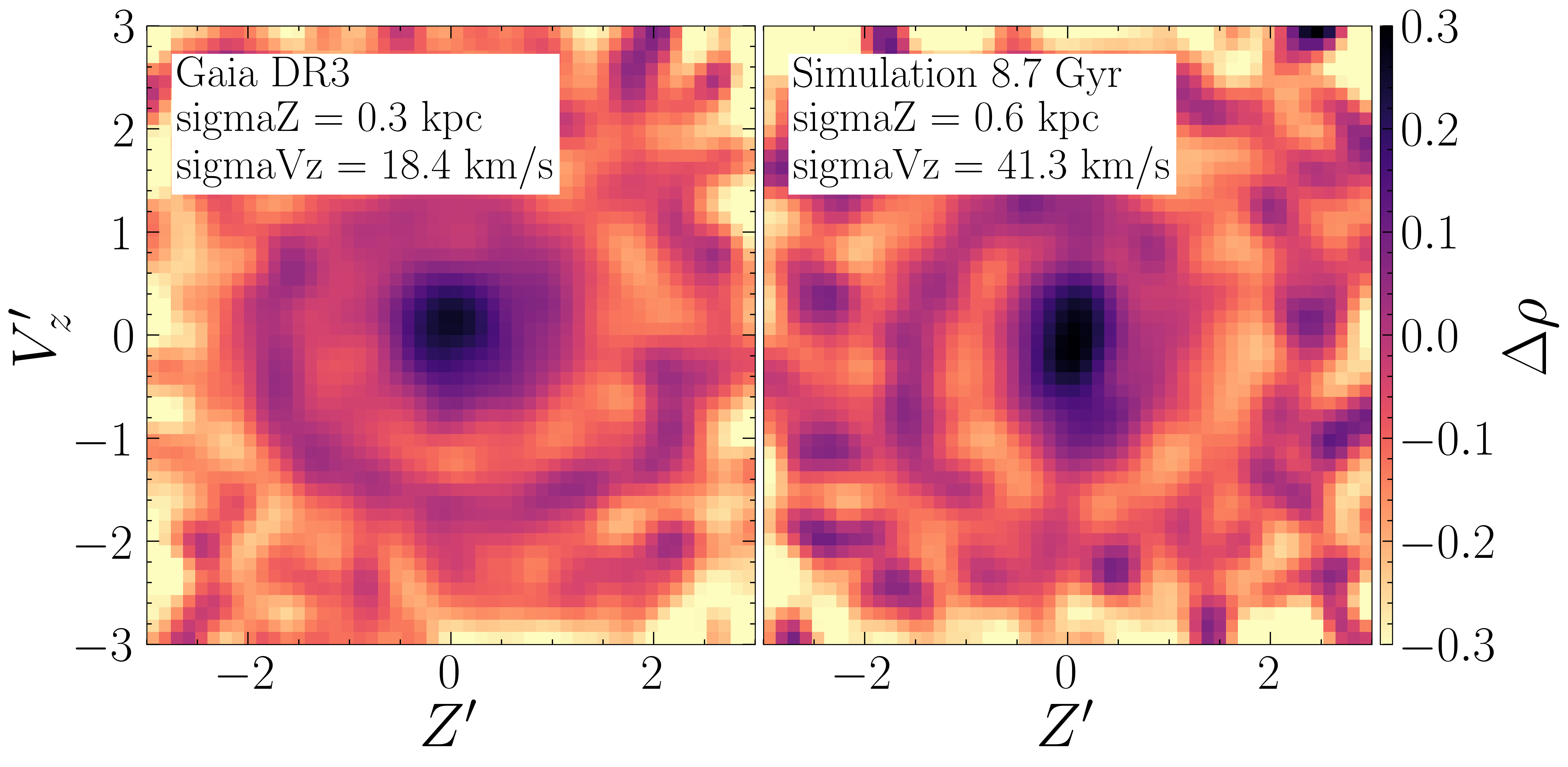}}
  \caption{The phase spiral in the \Gaia~DR3 data compared to that in an equivalent region in the warped galaxy at $8.7$~Gyr, color-coded by density contrast. The observational sample contains $\numprint{55880}$ stars. The simulation sample contains $\numprint{35989}$ star particles. Each axis shows the normalized phase space coordinate, covering the range $[-3, 3]$. The vertical position and velocity dispersions of each sample are indicated in the top left corner of each panel. }
  \label{fig:gaia_sim}
\end{figure}
To compare the amplitudes of the detected phase spirals in the warped simulation directly with \Gaia observations, we randomly select $\numprint{100000}$ stars within $1$~kpc of the Sun from \Gaia~DR3. We then select a subsample located within $\phi \in [150, 180]\degree$, which contains $\numprint{55880}$ stars. We transform the \Gaia observables into Galactocentric Cartesian coordinates, adopting the same position and velocity of the Sun as those in \citet{2023A&A...673A.115A}. In \figref{fig:gaia_sim}, we provide a direct comparison between the phase spiral in the observational sample and a qualitatively similar one within $R\in[7.277, 9.277]$~kpc and $\phi \in [30, 60]\degree$ of the warped simulation at $8.7$~Gyr. The simulation sample contains $\numprint{35989}$ star particles. The two phase space maps are color-coded by density contrast using the same color scale. \figrefalt{fig:gaia_sim} reveals that the phase spiral in the simulation sample has an amplitude comparable to that in the observational sample.

An idealized phase spiral model with a single perturbation time and no self-gravity would give rise to phase spirals whose winding decreases monotonically with radius \citep{2025arXiv250719579W}. The phase spirals observed in the high-resolution simulation of \citet{2025A&A...700A.109A} exhibit similar trends, where those in the inner disc appear to be more tightly-wound than those in the outer disc (see their Figs.~10 and 11). However, the observed phase spirals in \Gaia~DR3 display a relatively constant radial profile in the spiral winding \citep{2025arXiv250719579W}. In particular, there exist bands of high winding near the local spiral arms. The non-monotonic radial variation of the phase spiral winding has also been observed in the azimuthal action and angle space \citep[e.g.,][]{10.1093/mnrasl/slac082,10.1093/mnras/stad908}.

In this work, we find that the phase spiral winding in the warped simulation varies non-monotonically with guiding center radius (\figref{fig:rg_thetaphi_64_1450}), contrary to the prediction from the idealized model. The discrepancy between our results and the prediction is expected, because our phase spirals originate from multiple perturbations in a self-gravitating simulation. The details of the phase spiral winding in the warped simulation as a function of radius remains to be investigated. Intriguingly, we observe high winding from $R_g = 6.9$ to $9.1$~kpc (\figref{fig:rg_thetaphi_64_1450}), although we have not explored the correlation between the phase spiral winding and the location of the spiral arms.

\subsection{Impact time analysis}
As introduced in \secref{sec:intro}, the impact time of a single, impulsive perturbation can be inferred from the winding of the resulting phase spirals. Nevertheless, the formation mechanism proposed in this work involves multiple major perturbations. The response of the disc to multiple perturbations can be complicated. N-body simulations with multiple Sgr passages suggested that the most recent perturbation may `reset' the phase spiral signature \citep[e.g.,][]{2019MNRAS.485.3134L,2019MNRAS.486.1167B}, although \citet{10.1093/mnras/stac1491} did not observe a clear reset in their cosmological simulation. \citet{2025ApJ...988..254L} demonstrated that signatures of past perturbations may be preserved in the phase space. Most recently, \citet{10.1093/mnras/staf1463} proposed that the nonlinear coupling of two large-scale perturbations could induce the `Galactic echo' --- a second-order response which preserves the phase space imprint of earlier perturbations. If we consider the gas warp as one of the drivers of the observed phase spirals, it will further complicate the inference of a single impact time from an impulsive event such as the passage of the Sgr dwarf galaxy.

\subsection{Correlation between the phase spiral incidence rate and other properties}
\textit{Warp precession rate.}  \citet{2022MNRAS.512.3500K} measured a slow retrograde precession of the warp in the simulation, at a rate of $\sim -12 \kms \mathrm{kpc}^{-1}$. Their Fig.~9 shows the time evolution of the $m = 1$ bending signals in the warped simulation, measured at a time cadence of $\sim500$~Myr. In their Fig.~9, the warp precession rate is nearly constant with time, showing no detectable correlation with the phase spiral incidence rate in \figref{fig:incidence_rate_gas_inflow_flux}.

\textit{Star formation rate.}
The global star formation history of the warped simulation is provided in App.~\ref{sec:sfr}. In \figref{fig:sfr}, the global star formation rate displays an overall trend of decline with time, showing no clear correlation with the occurrence rate of the phase spirals in \figref{fig:incidence_rate_gas_inflow_flux}.

\subsection{Future prospects}

\textit{Phase spiral amplitude and winding.} Recently, \citet{2025arXiv250719579W} developed a method to fully characterize the morphology of phase spirals. In our future work this method could be used to conduct a detailed characterization of the warp-induced phase spirals, which could deepen the understanding of their formation and evolution. Quantitative analysis of the phase spiral amplitude in the warped simulation allows a direct comparison with the \Gaia observations. Analysis of the phase spiral winding provides information on the timing of perturbations across the disc.

\textit{Two-armed phase spirals.} 
Two-armed phase spirals are typically associated with breathing modes in the Galaxy \citep[see, e.g.,][]{10.1093/mnrasl/slac082, 2022MNRAS.511..784G,2023MNRAS.524.6331L,2024A&A...690A..15A,2025A&A...700A.109A}. \citet{10.1093/mnrasl/slac112} identified breathing motions in the warped and unwarped simulations used in this study, though we have not detected clear signatures of the two-armed phase spirals. With future higher resolution simulations, we hope to detect the signatures of two-armed phase spirals.

\textit{Macro-spiral.} Phase spirals in the warped simulation are widespread, a result of the major perturbation by the warp. \citet{10.1093/mnras/stab2580} and \citet{2024MNRAS.52711393H} demonstrated that large-scale perturbations of the disc could generate local phase spirals that collectively form a large-scale phase spiral (the `macro-spiral'). This signature may be key to differentiating between global perturbations and local stochastic kicks.

\textit{Perturbation signatures in the gas disc.}
The limited number of cold gas particles in the gas disc of the warped simulation renders it difficult to analyze their phase space signatures. However, the perturbation signatures in the gas disc remain an interesting future prospect.

\section{Conclusions}
\label{sec:conclude}
In this work, we have presented the first discovery of one-armed phase space spirals in the isolated, warped $N$-body$+$SPH simulation of \citet{2022MNRAS.512.3500K}. These phase spirals have been identified in the $Z - V_z$ map weighted by density contrast, residual azimuthal velocity and median radial velocity. The phase spiral pattern varies with metallicity, consistent with the observations \citep{2019MNRAS.486.1167B}. Viewing the detected phase spirals in the guiding center radius versus azimuthal phase angle plane demonstrates that they occur globally. The morphology of the phase spirals varies with location. Curiously, the phase spiral winding varies non-monotonically with guiding center radius.

Adopting a Fourier decomposition algorithm, we have been able to quantitatively track the time evolution of the one-armed phase spirals in the warped simulation. The incidence map, which illustrates the spatial and time variation of the emergence and strength of the phase spirals, demonstrates that the detected phase spirals are prevalent and long-lived.

We discovered that the incidence rate of the phase spirals evolves with the gas inflow at the outer disc and the bending wave amplitude. In particular, the phase spiral incidence rate lags behind the gas inflow by a few hundred Myr, consistent with the literature estimates for the delay between the time of the perturbation and the emergence of phase spirals \citep[see, e.g.,][]{10.1093/mnras/stab704,10.1093/mnras/stab2580,10.1093/mnras/staf1331,2025A&A...700A.109A}. There is no significant delay between the incidence rate and the bending wave amplitude. Hence, we propose that the detected one-armed phase spirals are signatures of the bending waves induced by irregular gas accretion along the warp. 
Repeated excitations by the warp sustain the phase spirals for $\sim 10$~Gyr, which implies that in this scenario there is no single perturbation time.

We have also detected phase spirals in the unwarped simulation. These phase spirals are weak and stochastically-distributed, their incidence rate displaying no strong correlation with the bending wave amplitude. They are likely induced by stochastic heating or the weak bending waves.

We have shown that the warp-induced phase spirals have consistent amplitudes with those in the \Gaia observations, although a more quantitative analysis is required.

A more quantitative analysis of the characteristics of the warp-induced phase spirals and higher resolution $N$-body$+$SPH simulations will allow us to characterize the phase spirals in more detail and provide further insight into the interpretation of the \Gaia observations.

\begin{acknowledgements}
We thank the referee for the insightful suggestions that improved this manuscript.
SW greatly appreciates the helpful discussions at the Lorentz Center workshop `Winding, Unwinding and Rewinding the \Gaia Phase Spiral'. SW is grateful to E.~Vasiliev for his insightful suggestions on the potential approximation and to T.~Asano for providing access to the simulation data from \cite{2025A&A...700A.109A}.

This project has received funding from the European Union's Horizon 2020 research and innovation programme under the Marie Sk{\l}odowska-Curie grant agreement No. 101072454 (MWGaiaDN).

The simulations in this paper were run at the DiRAC Shared Memory Processing System at the University of Cambridge, operated by the COSMOS Project at the Department of Applied Mathematics and Theoretical Physics on behalf of the STFC DiRAC HPC Facility (www.dirac.ac.uk). This equipment was funded by BIS National E-infrastructure capital grant ST/J005673/1, STFC capital grant ST/H008586/1, and STFC DiRAC Operations grant ST/K00333X/1. DiRAC is part of the National E-Infrastructure.

Preliminary analysis was carried out on Stardynamics, a 64- core machine that was funded from Newton Advanced Fellowship NA150272 awarded by the Royal Society and the Newton Fund.

This work has made use of data from the European Space Agency (ESA) mission {\it Gaia} (\url{https://www.cosmos.esa.int/gaia}), processed by the {\it Gaia} Data Processing and Analysis Consortium (DPAC, \url{https://www.cosmos.esa.int/web/gaia/dpac/consortium}). Funding for the DPAC has been provided by national institutions, in particular the institutions participating in the {\it Gaia} Multilateral Agreement.

This work made use of the following software: \code{astropy} \citep{astropy:2013,astropy:2018}, \code{matplotlib} \citep{matplotlib}, \code{scipy} \citep{scipy}, and \code{numpy} \citep{numpy}. The processing of the simulations was performed with the Python library \code{pynbody} \citep{pynbody}.
\end{acknowledgements}

\bibliographystyle{aa}
\bibliography{main}{}

\begin{appendix}

\section{Supplementary figures}

\subsection{Star formation history}
\label{sec:sfr}
\begin{figure}
  \resizebox{\hsize}{!}{\includegraphics{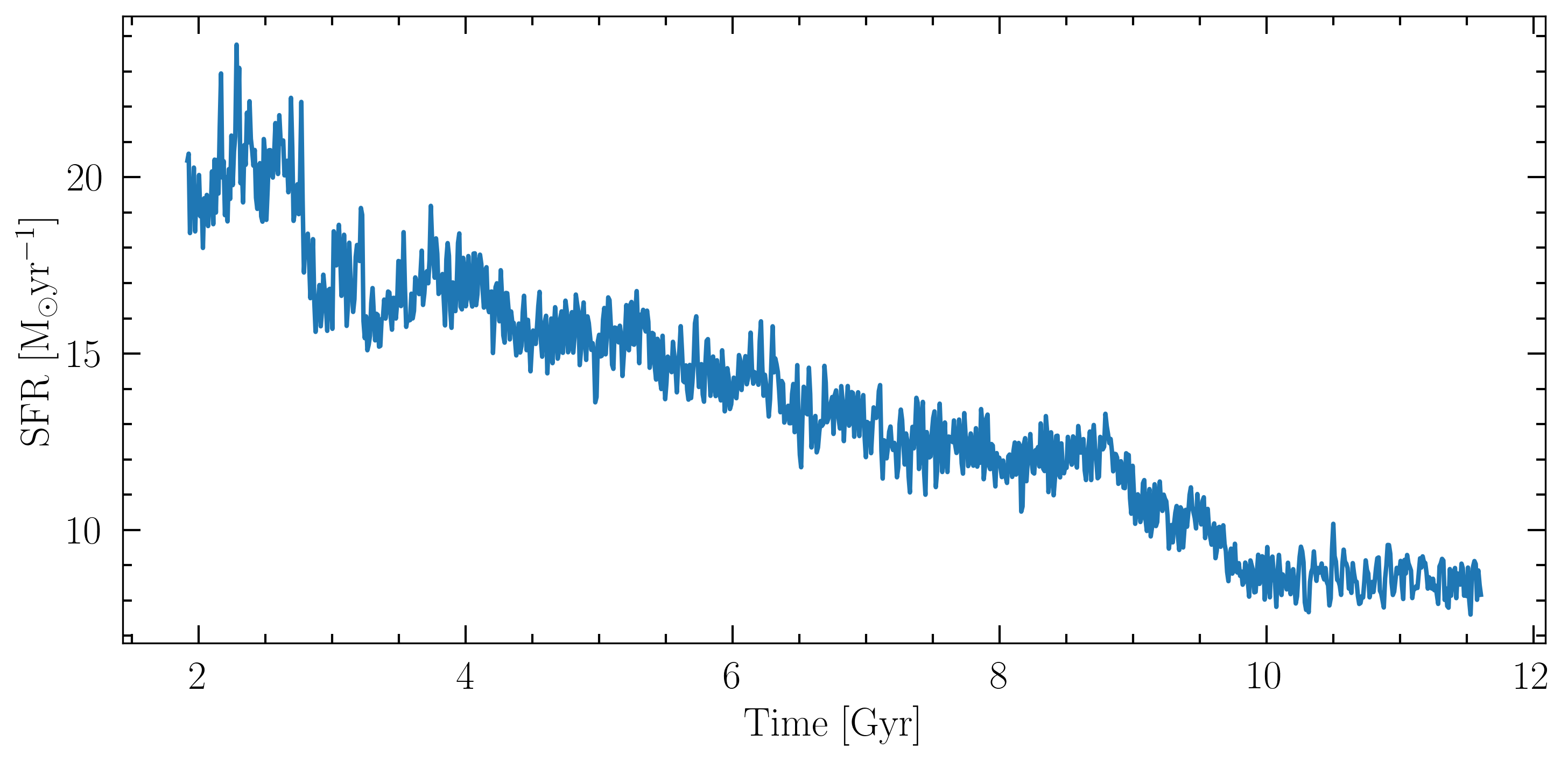}}
     \caption{Global star formation history of the warped simulation.}
     \label{fig:sfr}
\end{figure}

\figrefalt{fig:sfr} shows the global star formation history (SFH) of the warped simulation. The global star formation rate (SFR) displays an overall trend of decline with time. The SFR in \figref{fig:sfr} appears slightly higher than in the Milky Way. This is because the warped simulation in \citet{2022MNRAS.512.3500K} was run with a low density threshold for star formation in order to increase the number of star particles forming within the warp.
\FloatBarrier

\subsection{Velocity profiles}
\begin{figure*}
\sidecaption
  \includegraphics[width=12cm]{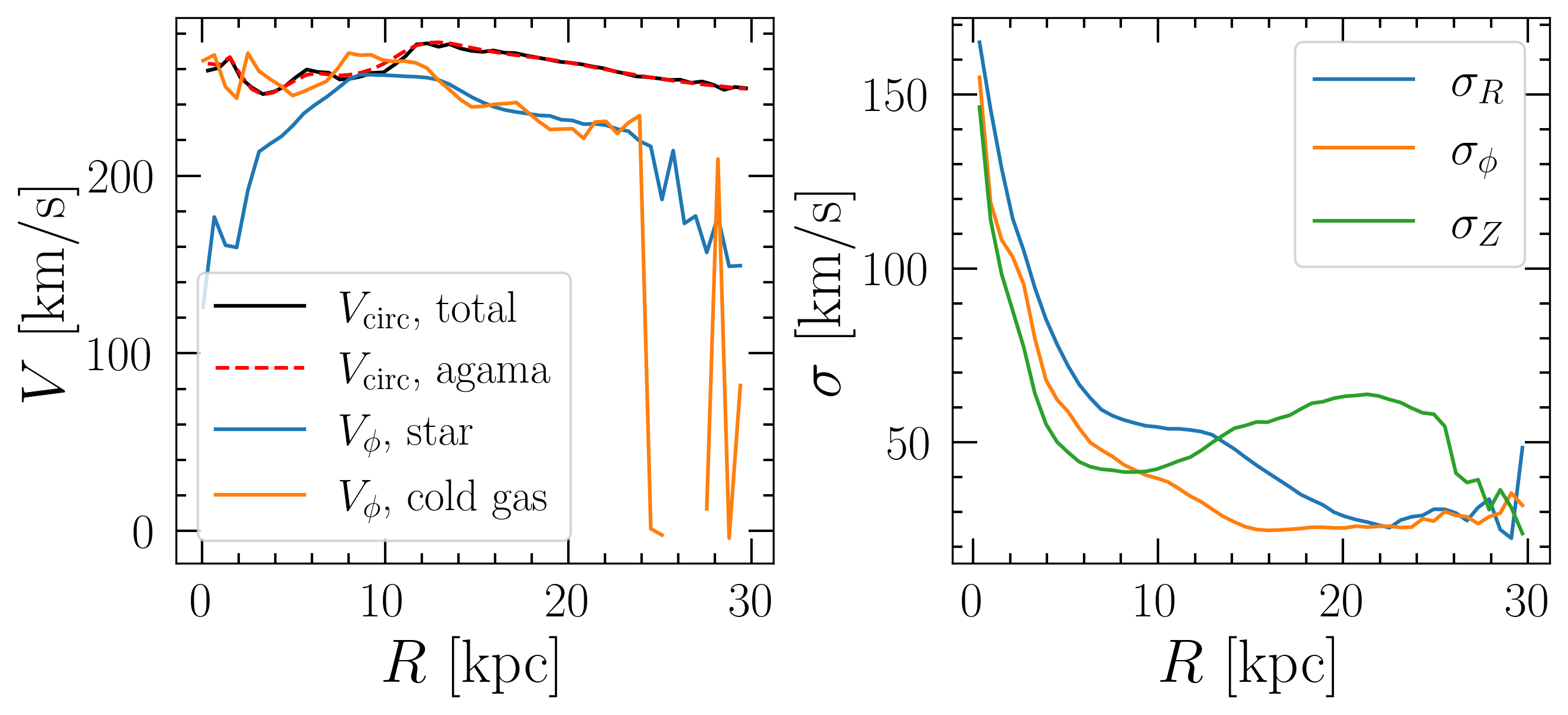}
     \caption{Left: the mass-averaged azimuthal velocity profile of the star particles and gas, as well as the circular velocity ($V_\mathrm{circ}$) profile for all particles and that calculated from the approximated potential using \code{agama} in the warped simulation at $8.7$~Gyr; Right: The mass-averaged radial ($\sigma_R$), azimuthal ($\sigma_{\phi}$) and vertical velocity dispersion ($\sigma_{Z}$) profiles of the same simulation snapshot.}
     \label{fig:1450_vel_disp_rot_vel}
\end{figure*}

The left panel of \figref{fig:1450_vel_disp_rot_vel} shows the mass-averaged azimuthal velocity profile of the star particles and cold gas ($T < \numprint{50000}$~K), as well as the circular velocity ($V_\mathrm{circ}$) profile for all particles (star particles, gas and dark matter) and that calculated from the approximated potential using \code{agama} in the warped simulation at $8.7$~Gyr. The right panel shows the radial ($\sigma_R$), azimuthal ($\sigma_{\phi}$) and vertical velocity dispersion ($\sigma_{Z}$) profiles of the same simulation snapshot.

\subsection{Phase spirals in the $R_g$-$ \theta_{\phi}$ plane (high $J_R$ population)}
\begin{figure*}
\centering
   \includegraphics[width=17cm]{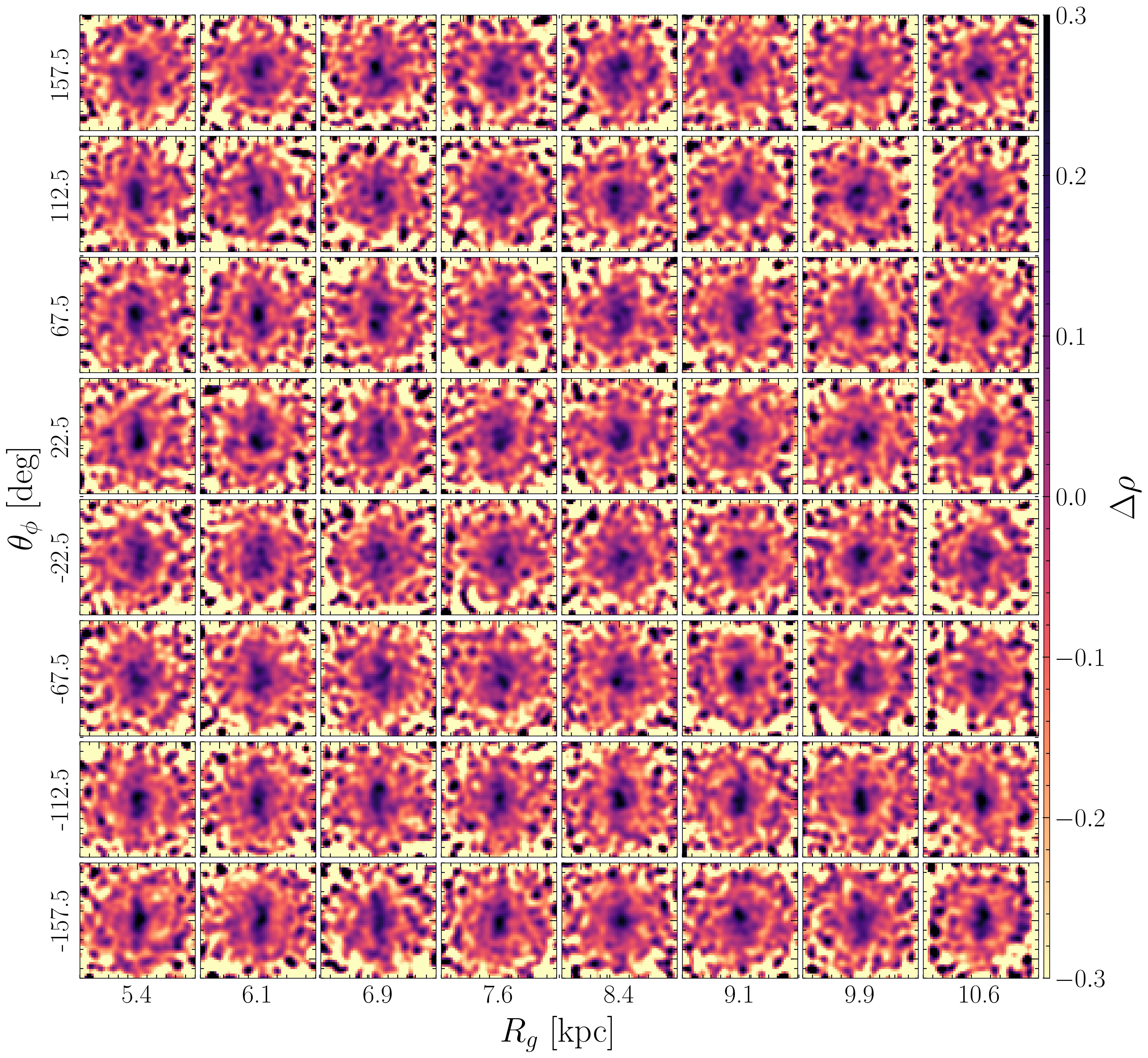}
     \caption{The phase spirals in the $R_g$-$ \theta_{\phi}$ plane for star particles with $J_R>\mathrm{median}(J_R)$ at $8.7$~Gyr in the warped galaxy, color-coded by density contrast.}
     \label{fig:rg_thetaphi_64_1450_highjr}
\end{figure*}

\figrefalt{fig:rg_thetaphi_64_1450_highjr} shows the detected phase spirals in the $R_g$-$ \theta_{\phi}$ plane for star particles with high radial action. These star particles are dynamically old, exhibit weaker bending waves \citep{2022MNRAS.512.3500K} and less prominent phase spiral signatures.

\subsection{Examples of the Fourier decomposition algorithm}
\label{sec:fourier}

We use analytical one-armed phase spirals to validate the phase spiral finder algorithm. We create a $100\times 100$ grid in the vertical phase space with $Z \in [-3, 3]$~kpc and $V_z \in [-120, 120] \kms$. We add Gaussian noise to simulate the vertical position and velocity dispersions, with $\sigma_Z = 1$~kpc and $\sigma_{V_z} = 60 \kms$, respectively. The analytical expression simulating a one-armed phase spiral in the mock $\Delta V_{\phi}$-weighted phase space is:
\begin{equation}
    \Delta V_{\phi}(\tilde{R}, \tilde{\phi}) = A_1(\tilde{R})  \sin(\tilde{\phi} - P_1(\tilde{R})) \,,
    \label{Eq:1_armed_model}
\end{equation}
where $A_1$ and $P_1$ linearly increase with $\tilde{R}$. After generating the analytical phase spiral, we apply the phase spiral finder algorithm and Fourier decompose the normalized phase space diagram.

\begin{figure*}[h!]
    \centering
    {\includegraphics[clip,scale=0.4] {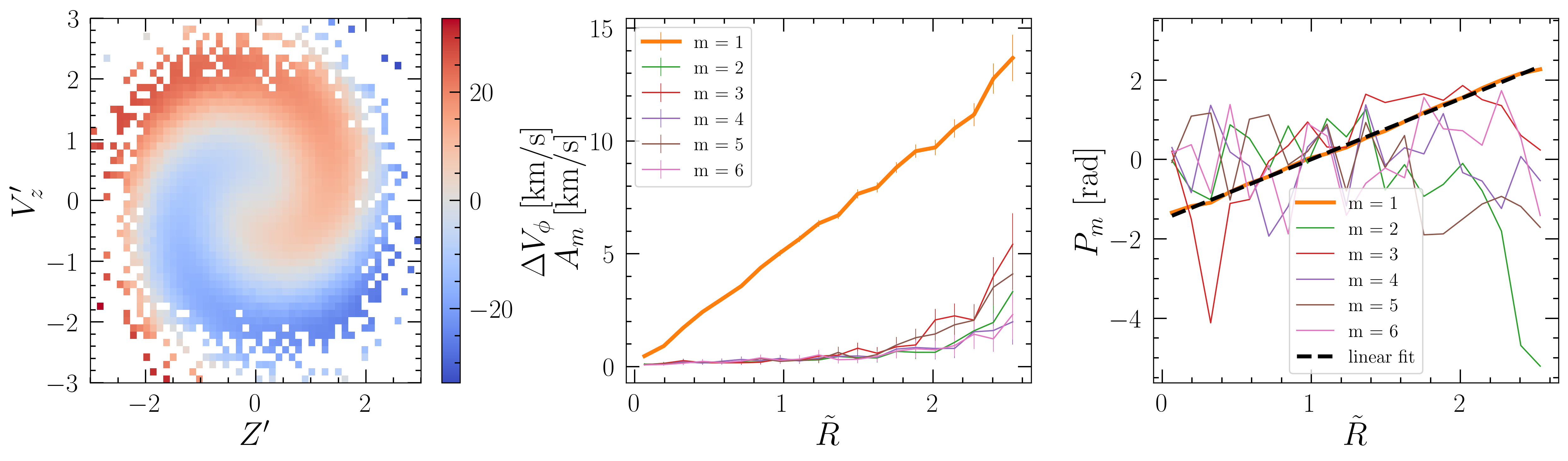}}
     \caption{Left: mock $\Delta V_{\phi}$-weighted phase space map for an idealized one-armed phase spiral. Middle and right: the result of Fourier decomposing the phase space map . Middle: Fourier amplitude versus phase space radius with $m = 1, \cdots, 6$. Error bars are calculated by bootstrapping the sample of star particles within each annulus $200$ times. Right: Fourier phase angle versus phase space radius. The black dashed line shows a linear fit to the $m = 1$ $P_1$-$\tilde{R}$ curve. }
      \label{fig:model_fourier_decomp}
\end{figure*}

\figrefalt{fig:model_fourier_decomp} shows the result of Fourier decomposing the analytical one-armed phase spiral with $A_1(\tilde{R}) = -10 \tilde{R}$ and $P_1(\tilde{R}) = 0.5 \pi \tilde{R}$. The plot demonstrates the effectiveness of the algorithm in the identification of one-armed phase spirals.

\begin{figure*}[h!]
    \centering
    {\includegraphics[clip,scale=0.4] {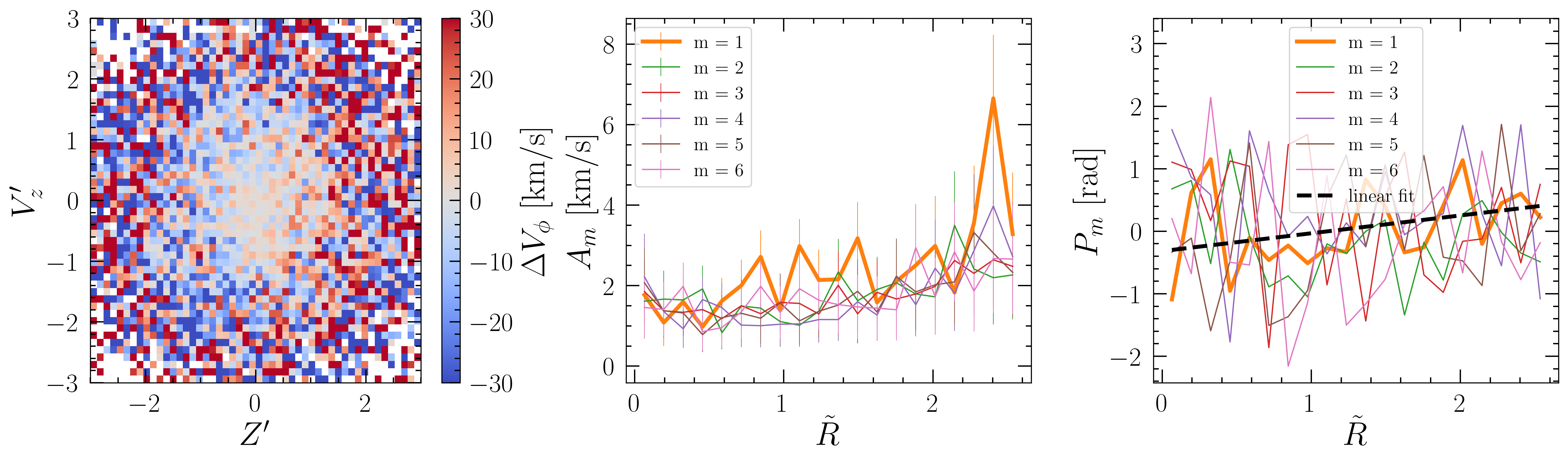}}
     \caption{Left: $\Delta V_{\phi}$-weighted phase space diagram in region~4 of the warped galaxy at $8.7$~Gyr. Middle: the radial profile of the Fourier amplitude, for $m = 1, \cdots, 6$. Right: the radial profile of the Fourier phase angle. The black dashed line shows a linear fit to the $m = 1$ $P_1$-$\tilde{R}$ curve. }
      \label{fig:1450_spiral_fourier_decomp}
\end{figure*}

\figrefalt{fig:1450_spiral_fourier_decomp} shows the result of Fourier decomposing the $\Delta V_{\phi}$-weighted phase space map in \figref{fig:vzz_dens_vphi_vr} (region~4 in the warped galaxy at $8.7$~Gyr). The dominant $A_1$ and linearly increasing $\phi_1$ correspond to a positive detection of a one-armed phase spiral.

\onecolumn
\end{appendix}

\end{CJK*}
\end{document}